\documentclass[prd,showpacs,preprintnumbers,nofootinbib,aps]{revtex4}
\usepackage{subeqnarray}
\usepackage{epsfig,amsmath,amssymb}
\usepackage{mathrsfs}
\usepackage[usenames,dvipsnames]{color}
\usepackage[pagebackref=true, colorlinks=true]{hyperref}
\definecolor{redish}{rgb}{0.7,0.2,0.0}  
\definecolor{bluish}{rgb}{0.2,0.5,0.8}
\hypersetup{linkcolor=redish,          
                  citecolor=blue,        
                  filecolor=magenta,      
                  urlcolor=bluish}          
\DeclareFontFamily{U}{rsfs}{}         
\DeclareFontShape{U}{rsfs}{m}{n}{<5> rsfs5 <6><7> rsfs7          %
  <8><9><10><10.95><12><14.4><17.28><20.74><24.88> rsfs10}{}     %
\DeclareMathAlphabet{\mathfs}{U}{rsfs}{m}{n}                     %
\newcommand{\mfs}[1]{\mathfs {#1}}                               %

\newcommand{\ba}{\nopagebreak[3]\begin{eqnarray}}
\newcommand{\ea}{\end{eqnarray}}
\newcommand{\bii}{\begin{itemize}}
\newcommand{\eii}{\end{itemize}}
\newcommand{\nn}{\nonumber}

\newcommand{\sO}{{\mfs O}}

\newcommand{\f}{\frac}
\def \p{\partial}
\def \d{\delta}
\def \b{\beta}
\def \l{\ell}
\def \g{\gamma}

\def \lp{\l_p}
\def \j{\sqrt{j(j+1)}}

\def \lm{\lambda}
\def \a{\approx}
\def \s{\sigma}
\def \n{\eta}

\def \H{\mathcal{H}}

\def \sj{s_j^{\star}}

\def \cs{\{s_j\}}
\def \bcs{\{\sj\}}

\def \E{E^{\star}}
\def \A{A^{\star}}
\def \Z{Z_C}
\def \Zs{\Z^{\star}}
\def \O{\Omega}
\def \o{\omega}

\def \L{\Lambda}
\def \h{\hat}

\def \({\left(}
\def \){\right)}
\def \[{\left[}
\def \]{\right]}
\begin{document}

\title{Thermodynamics of Quantum Isolated Horizons with  model Hamiltonians }
\author{Abhishek Majhi}%
 \email{abhishek.majhi@saha.ac.in}
\affiliation{Astro-Particle Physics and Cosmology Division\\Saha Institute of Nuclear Physics\\Kolkata-700064, India}%

\begin{abstract}
Following a recent proposal, we consider the most general structure possible for the Hamiltonian operator associated with the Quantum Isolated Horizon(QIH) with explanations of the underlying physical motivations. An extensive thermodynamic analysis with this model Hamiltonian is presented.  Considering {\it fixed} number of punctures,  the canonical partition function can now be written in the usual energy ensemble. Arguing that the known classical results must follow in the correspondence limit (viz. the equilibrium temperature as observed by a local observer) the model is fixed, yielding the energy spectrum of the QIH.  
\end{abstract}
\maketitle
\section{Introduction}
The classical {\it Isolated Horizon}({\bf IH}) is a null inner boundary of spacetime, foliated by 2-spheres, with specific local properties consistent with General Theory of Relativity and is a generalization of the teleological concept of event horizons to more realistic and dynamical situations that are expected to occur in Nature \cite{ih1,ih2,ih3,ih4,ih5}. The quantization of  IH phase space \cite{qg1,qg2} in the {\it Loop Quantum Gravity}({\bf LQG}) framework provides the topology of {\it Quantum Isolated Horizon}({\bf QIH}), at a particular time slice, to be that of 2-sphere punctured by the spin network describing the bulk quantum geometry. The QIH degrees of freedom belong to the Hilbert space of an $SU(2)$ {\it Chern-Simons}({\bf CS}) theory coupled to the punctures, acting as sources. Recently it has been shown \cite{me1,me2} that the microcanonical entropy of a QIH can be completely  written in terms of the two macroscopic parameters, $k$ and $N$, which fully characterizes the macrostates of the QIH, applying standard statistical mechanical methods and using the knowledge of the physical degrees of freedom of the QIH belonging to the Hilbert space of the CS theory. To mention, $k$ is the level of the CS theory and $N$ is the total number of punctures on the QIH. It has been also shown that the {\it Barbero-Immirzi}({\bf BI}) parameter$(\g)$ \cite{bar1,bar2,im1,im2} must be bounded within a certain range of values for the {\it Bekenstein-Hawking area law}({\bf BHAL})\cite{bhal} to be valid. However, by studying a system in the microcanonical ensemble we only get a statistical viewpoint. The true thermodynamic analysis of a system begins only in the canonical or grand canonical ensemble where thermal fluctuations are allowed, although they are considered to be negligible in the context of equilibrium thermodynamics. But, to write down the canonical partition function, we need to have the knowledge about the Hamiltonian associated with the QIH. Unfortunately, there is none in the present literature. Although there is a well established notion of classical energy associated with the IH which obeys first law \cite{ih3}, but in the quantum theory there is no known Hamiltonian operator or any energy spectrum associated with a QIH resulting from a true quantization of the classical theory, which, hopefully will be done someday. 
In this work, we do not attempt to do this quantization. Instead, based upon some  well motivated physical arguments we propose the most general structure of the Hamiltonian that a QIH can have  and perform an extensive thermodynamic analysis. Considering the model Hamiltonian to yield the known classical results of  thermodynamics, the unknown coefficients are suitably chosen and hence we obtain the Hamiltonian operator and the corresponding energy spectrum of a QIH.
\par
Here is an outline of the subject matter of this paper. In section(\ref{sec2}), we briefly review and discuss the structure of the model Hamiltonian, hence the corresponding energy spectrum, for QIH proposed in a companion paper \cite{hamqih}, based on the knowledge of the theory of QIH and area operator in LQG framework. 
The structure of the Hamiltonian ensures that it is gauge invariant, self-adjoint and commutes with the area operator implying that the mean area of the QIH is a constant of motion, which signifies that the constant classical area property of IH is emergent. In section(\ref{sec3}), we lay down an extensive thermodynamic analysis with the model Hamiltonian. Having an explicit Hamiltonian for the QIH at our disposal, we are able to write the canonical partition function for quantum spacetimes admitting QIH as an inner boundary considering the usual {\it energy} ensemble.
We construct the canonical partition function for a quantum spacetime admitting a QIH as its inner boundary by the thermal holographic approach introduced in \cite{pm1} and generalized for the grand canonical ensemble in \cite{ampm1}. We perform a rigorous thermodynamic analysis. The expressions for the equilibrium temperature being plagued with the unknown coefficients of the model, we make suitable choices of these coefficients so as to match the expression of equilibrium temperature with the known results of the classical theory. Firstly, in section(\ref{sec3.1}), a simplified model of the energy spectrum is studied where the single puncture energy contribution is related to the single puncture area contribution by a  power law only, which helps us fixing the coefficients. In section(\ref{sec3.2}), using the determined coefficients back into the general energy spectrum the analysis is completed. As a result we obtain the expression for the equilibrium temperature and finally the general Hamiltonian operator and corresponding energy spectrum of the QIH. Finally, we conclude with a discussion in section(\ref{sec5}) explaining the relevance of this work as far as the  literature of black hole thermodynamics is concerned.

\section{Model Hamiltonian and Energy spectrum of QIH}\label{sec2}

Even though the first law associated with an IH\cite{ih3} gives us the notion of a well defined classical energy associated with the horizon, there has never been an attempt to deal with the energy spectrum of a QIH until recently in \cite{gp}, where, based on some ad hoc arguments and semiclassical approximations it has been proposed that the area spectrum of the QIH, with some scale factor is itself the energy spectrum of the QIH. However, the true quantization of the horizon energy is still lacking.
In this work, we shall not deal with the quantization of the classical energy. Instead we shall take the quantum theory as the starting point and consider a model Hamiltonian operator for the QIH, hence obtaining a well defined energy spectrum for the same, proposed in a companion paper\cite{hamqih}. The proposed model will be fixed so as to match the known classical results. In this section, we shall review in brief and discuss the relevant details and physical motivations behind considering the model Hamiltonian proposed in \cite{hamqih}.

\subsection{The area operator for QIH}
As we begin with the quantum theory, the available knowledge at our disposal and relevant in this context are the quantum geometric area operators and area spectrum in LQG\cite{area1,area2} and the theory of QIH\cite{qg1,qg2}. The area operator is a gauge invariant, self-adjoint observable in LQG defined for any arbitrary two dimensional surface $(S)$ embedded in the three dimensional spatial manifold $(\Sigma)$ obtained from a specific foliation of the four dimensional spacetime manifold $(M\equiv R\times\Sigma)$ by some preferred time evolution vector field \cite{area1,area2}. The area operator and the corresponding spectrum for a QIH is completely known to us. For a given number of punctures $(N)$, the Hilbert space structure of the QIH is given by Inv$(\bigotimes_{l=1}^N\H_{j_l})$, where $j_{l}$ is the spin at the $l$-th puncture and $1/2\leq j_l\leq k/2, \forall l\in[1,N]$, $k$ being the level of the associated CS theory\cite{qg1,qg2,wit} and `Inv' stands for gauge invariance. For a given spin sequence, the area spectrum is given by
\ba
\hat A|j_1,\cdots,j_N\rangle=8\pi\g\lp^2\sum_{l=1}^N\sqrt{j_l(j_l+1)}|j_1,\cdots,j_N\rangle
\ea
It should be noted that the area spectrum of the QIH is different from that of any arbitrary surface due to the fact that, unlike for a QIH, for an arbitrary surface the spins can have values $0,1/2,1,\cdots, \infty$ i.e. the spectrum is unbounded above and also the lower bound is different from that of a QIH. The structure of the Hilbert space of the QIH  prompts us to write down the area operator as
\ba
\h A\equiv \h A_{j_1}\otimes \h I_{j_2}\otimes\cdots\otimes\h I_{j_N}+\h I_{j_1}\otimes \h A_{j_2}\otimes\cdots\otimes \h I_{j_N}+\cdots +\h I_{j_1}\otimes \h I_{j_2}\otimes\cdots\otimes \h A_{j_N}\label{arpun}
\ea
Looking at the description of a QIH in the LQG framework, it is quite explicit that the punctures i.e. the intersection points of the bulk spin network with the IH, are the elementary quantum building blocks of the QIH \cite{qg1,qg2}. It is supported by the  fact that all the essential properties  of the QIH are manifested at the punctures.   The connections are flat everywhere on the horizon except at the punctures. The holonomies of the connection are nontrivial {\it only} along the disjoint  loops enclosing the punctures, otherwise trivially equal to unity. Last but not the least, all the essential and relevant properties of the area operator (e.g. self adjointness, gauge invariance,etc.) are indeed manifested by these individual punctures or intersection points\cite{area1,area2}. This particular way of looking at issues related to QIH is due to \cite{hamqih} and one needs to look into it to understand the motivations behind this viewpoint of attacking the problem initially from a quantum perspective.

\subsection{Structure of the Hamiltonian operator and its properties}
Now, the scenario at once compels us to think of any operator belonging to the QIH Hilbert space as being contributed from the individual punctures and should have a structure like that of the area operator in (\ref{arpun}). Hence, it follows that the Hamiltonian operator for the QIH should be written as
\ba
\h H_S\equiv \h H_{j_1}\otimes \h I_{j_2}\otimes\cdots\otimes\h I_{j_N}+\h I_{j_1}\otimes \h H_{j_2}\otimes\cdots\otimes \h I_{j_N}+\cdots +\h I_{j_1}\otimes \h I_{j_2}\otimes\cdots\otimes \h H_{j_N}
\ea
The area operator is a gauge invariant observable follows from the fact that when the bulk spin network pierces the surface $S$, the puncture is assigned with an $SU(2)$ spin and the area contribution from that puncture is nothing but the Casimir of the $SU(2)$ gauge group of the underlying theory \cite{area1,area2}. Hence, we can argue that the `powers' of the Casimir being also gauge invariant can as well be the contribution from a single puncture to some other gauge invariant observable smeared over the QIH, say the Hamiltonian. So, we propose that the most general form of the contribution of the Hamiltonian from a single puncture to be of the form
\ba
\h H_j\equiv \lp\sum_{n=0}^{\L}b_n \hat A_j^n\label{hamsin}
\ea 
where $\L$ is a necessary cut-off and $b_n$ are coefficients with proper dimension (to be discussed shortly). Hence, the Hamiltonian operator for the QIH can now be written as 
\ba
\h H_S\equiv\lp\sum_{n=0}^{\L}b_n \left(\h A_{j_1}^n\otimes \h I_{j_2}\otimes\cdots\otimes\h I_{j_N}+\h I_{j_1}\otimes \h A_{j_2}^n\otimes\cdots\otimes \h I_{j_N}+\cdots +\h I_{j_1}\otimes \h I_{j_2}\otimes\cdots\otimes \h A_{j_N}^n\right)
\ea
Besides the gauge invariance of the above operator, there is yet another property which follows from its construction. It is straightforward to see that it commutes with the area operator 
\ba
\left[\h H_S ,\h A\right]
&\equiv&\h 0\label{commu}
\ea
The crucial implication that it has, due to this commutativity, is that, if there is some evolution parameter $\tau$ with respect to which the QIH evolves, then it is evident that
\ba
\f{d}{d\tau}\langle\h A\rangle=\f{1}{i\hbar}\langle\left[\h H_S ,\h A\right]\rangle=0
\ea
i.e. the expectation value of the area operator of a QIH is a constant of motion. This is perfectly consistent with the fact that in the correspondence limit, the most crucial classical property of the corresponding classical Isolated Horizon i.e. fixed classical area, emerges as a consequence of the construction of the model Hamiltonian. Further, as a result of this commutativity in eq.(\ref{commu}), we can also say that the states of the quantum CS theory describing the local quantum degrees of freedom of the QIH, are simultaneous eigenstates of the area operator and the model Hamiltonian for the QIH. 

Thus, we have successfully justified the proposal of the model Hamiltonian for the QIH based on strong {\it physical motivations} which can be briefly restated as follows :-
\begin{itemize}
\item The punctures are the most fundamental and elementary constituents of the QIH which collectively provide an effective description of the IH in the correspondence limit. 
\item The model Hamiltonian shares all the necessary and relevant properties of the area operator e.g. gauge-invariance, self-adjointness, etc.
\item The model Hamiltonian and the area operator associated with the QIH have simultaneous eigenstates which are those of the CS theory coupled to punctures.
\item The structure of the model Hamiltonian ensures that the constant area property of IH emerges in the correspondence limit.
\end{itemize}

\subsection{Spectrum of the Hamiltonian operator}
Now, from  (\ref{hamsin}), the single puncture contribution to the energy spectrum can be written as
\ba
E_j=\lp\sum_{n=0}^{\L}b_nA_j^{~n}=\lp\sum_{n=0}^{\L}a_n(\g)C_j^{~n}\label{ej}
\ea
where $\text{Dim}[b_n]=\lp^{-2n}$ and $a_n(\g)=b_n(8\pi\g\lp^{2})^n$. It follows that the spectrum of a QIH with $N$ punctures looks like
\ba
\hat H_S|j_1,\cdots,j_N\rangle=\lp\sum_{l=1}^N\sum_{n=0}^{\L}a_n(\g)C_{j_l}^{~n}|j_1,\cdots,j_N\rangle
\ea
The above form of the energy spectrum spectrum has been written just for the sake of clarity. We shall work in the spin configuration basis in what follows, as it will be convenient for our calculations. For simplicity, in our model, we consider $n$ to take only integral values and $n\geq 0$ to ensure incremental monotonicity of the energy spectrum with the area contribution of a single puncture. Since this kind of model Hamiltonian or energy spectrum of the QIH has not been studied previously in literature and any property of such spectrum is hitherto unknown, to avoid any problem with the convergence of such spectrum we have used the cut-off parameter$(\L)$ {\it a priori}. We shall see that, at least from the thermodynamic viewpoint, we can assert that such a cut-off is indeed required and should emerge automatically from a true quantization of the horizon energy, if can be done anyhow. This is because, as we proceed, the parameter $\L$ will come out to be directly related to the equilibrium temperature of the QIH and one does not expect it to diverge for thermodynamically stable systems.

\par
Of course there remain several questions to be answered about this model Hamiltonian as far as its candidature of being the true Hamiltonian of the QIH, which will result from a true quantization of the theory, is concerned, resolving which is itself a mammoth task to complete with various technical subtleties to overcome. As far as this work is concerned, our aim is to analyze the thermodynamic properties of the QIH considering the most general structure of the Hamiltonian one can construct from a pure quantum viewpoint devoid of any classical notions. Although the physical motivations behind the proposal of the model Hamiltonian has been discussed in details but the motivations behind the approach from quantum to classical rather than trying to quantize the classical theory is beyond the scope of this paper and are worth having separate explanations. To have a taste of the motivations behind adopting this particular viewpoint, which is meant for tackling other problems besides the issue of the Hamiltonian, as far as the theories of IH and QIH are concerned,  one can see \cite{hamqih}.

\section{Thermodynamic Analysis with the Hamiltonian}\label{sec3}

\subsection{The Canonical partition function}
The obvious consequence of the knowledge about the Hamiltonian of a system is the ability to write down the canonical partition function leading to the corresponding  thermodynamic analysis. This is what we are going to explore in what follows.  
Following \cite{pm1,ampm1} the canonical partition function for spacetimes admitting QIH as its internal boundary can be written as
\ba
Z_C= Tr(\exp -\b \hat H)\label{zc1}
\ea
where we have considered an ensemble of spacetimes, admitting QIH as an inner boundary, in contact with a heat bath with inverse temperature $\b$. $\hat H$ is the Hamiltonian for the system. Now, unlike the classical theory, the degrees of freedom of the quantum theory are independent as far as the bulk and the boundary are concerned \cite{qg1,qg2}. The Hilbert space for the quantum spacetime, with QIH as the inner boundary$(S)$ of the bulk quantum geometry$(B)$, can be written as $\H=\H_S\otimes\H_B$. As a result we can write the Hamiltonian for the thermodynamic system as 
\ba
\hat H\equiv\hat H_S\otimes\hat I_B + \hat I_S\otimes\hat H_B\label{hbs}
\ea
where $\hat I$ are the identity operators for the corresponding Hilbert spaces designated by the suffixes. Similarly, the wave function for a quantum spacetime, admitting QIH as an internal boundary, can be written as 
\ba
|\Psi\rangle=\sum_{S,B}C_{SB}|\Psi_S\rangle\otimes|\Psi_B\rangle\label{psi}
\ea 
It should be noted that the Hilbert space, the quantum states, etc. are all results of the kinematic quantization of classical degrees of freedom in a suitable choice of time slicing. Now, from the knowledge of the Hamiltonian formulation of General Relativity(GR) we know that, GR being a covariant theory, the bulk Hamiltonian is zero. Hence, we can write the quantum version of this constraint as 
\ba
\hat H_B|\Psi_B\rangle\a 0\label{qhc}
\ea
Rewriting the partition function in eq.(\ref{zc1}) in terms of the bulk and the boundary sectors using eq.(\ref{hbs}), eq.(\ref{psi}) and applying the quantum Hamiltonian constraint given by eq.(\ref{qhc}) we obtain
\ba
Z_C&=& \sum_{S,B}|C_{SB}|^2\langle\Psi_B|\otimes\langle\Psi_S|\exp -\b (\hat H_S\otimes\hat I_B + \hat I_S\otimes\hat H_B)|\Psi_S\rangle\otimes|\Psi_B\rangle\nn\\
&=&\sum_{S}|\bar C_{S}|^2\langle\Psi_S|\exp -\b\hat H_S|\Psi_S\rangle\nn
\ea
where $|\bar C_{S}|^2=\sum_B|C_{SB}|^2\langle\Psi_B|\Psi_B\rangle$. Now, one should {\it not} consider this QIH Hamiltonian$(\hat H_S)$ to be that of the quantum CS theory. Being a {\it topological} field theory, the Hamiltonian of the CS theory vanishes. This would have resulted in the partition function to be zero ! On the other hand there is another Hamiltonian defined on the IH which provides a well defined notion of classical energy and satisfies the first law under the Hamiltonian evolution of space in time{\cite{ih3}}. But, till now this notion of energy has no quantum version.
One can see \cite{hamqih} for a discussion on this issue and the prediction of a possible way to handle this problematic situation i.e. the quantum to classical viewpoint. The proposal of the model Hamiltonian, which we shall use here as $\hat H_S$ is an outcome of that line of thought. The eigenvalues of this Hamiltonian will give us the quantum energy spectrum of the QIH and the eigenvalue equation can be written as
\ba
\hat H_S|\Psi_S\rangle=E_S|\Psi_S\rangle
\ea 
Hence, the partition function can be simply written as
\ba
Z_C&=&\sum_{S}|\bar C_{S}|^2\exp -\b E_S\label{zcs}
\ea
In eq.(\ref{zcs}), $|\bar C_{S}|^2$ denotes the probability of finding the QIH at a particular surface state irrespective of the states of the bulk quantum geometry. Now, we do not have a quantum version of the Hamiltonian of the horizon introduced in \cite{ih3}. But in the previous section we have argued that similar to the area spectrum of the QIH in the LQG framework, we can construct an energy spectrum of the QIH by virtue of the additive contributions of the noninteracting punctures. Thus, using the energy spectrum given by eq.(\ref{espec}) and rewriting the partition function as sum over spin configuration we get 
\ba
\Z&=&\sum_{\cs}~\O[\cs]~\exp -\b E_{\cs}\nn
\ea
The state having energy $E_{\cs}$ exhibits $\O[\cs]$-fold degeneracy which in the continuum limit can be viewed to be the density of states. Since, in equilibrium statistical mechanics, the number of microstates $\O[\{\sj\}]$ corresponding to the most probable configuration $\sj$ is very very large compared to the number of microstates for other less probable subdominant configurations, we can write the exact partition function as the sum of the dominant part (equilibrium contribution) and the sub-dominant part (contributions from fluctuations about equilibrium) as follows 
\ba
Z_C&=&\O[\bcs] \exp -\b E_{\bcs}+~~\text{sub-dominant terms}\nn\\
&=&\O[\bcs] \exp -\b \E +~\bar\d\nn\\
&=&\Zs+~\bar\d\label{zcd}
\ea
where $E_{\bcs}=\sum_j\sj E_j=\E$ and $\bar\d$ denotes the contributions from the sub-dominant configurations contributing to the thermal fluctuations. Having calculated the partition function it is straightforward to show that the ensemble average of the energy of the QIH comes out to be
\ba
\bar E=-\f{\p}{\p\b}\log Z_C\simeq\E
\ea
Now, taking logarithm of both sides of eq.(\ref{zcd}) and using the definition of canonical entropy $S_C=\log Z_C+\b \bar E$ and using $\bar E\simeq\E$  from eq.(\ref{exe}) it is straightforward to show that
\ba
S_C=S_{MC}+\log(1+\bar\d/\Zs)
\ea
This is a familiar result previously shown in literature in the context of black hole thermodynamics in \cite{pm1,stab1,stab2,stab3,stab4,stab5,stab6,stab7}, but in a different mathematical approach where the fluctuations appear as terms of Taylor expansion about the equilibrium (see \cite{landau}).

\subsection{Entropy}
In equilibrium statistical mechanics related to thermodynamics of a system in general, working in the canonical ensemble is tantamount to working in the microcanonical ensemble as long as the fluctuations are small and can be ignored. The above analysis quite clearly shows that the same is true in the present scenario related to QIH thermodynamics. Since the basic calculations regarding the QIH thermodynamics in microcanonical ensemble are already available in literature, we shall use those here.
Recently it has been argued in \cite{me2} that taking into account all possible spins on the QIH, the CS level $(k)$ and the total number of punctures $(N)$ are the only macroscopic parameters which characterize the macrostates of the QIH. Thus, defining the microcanonical ensemble for given $k$ and $N$, the microcanonical entropy of QIH in terms of these two macroscopic parameters is given by\cite{me1,me2}
\ba
S_{MC}=\f{\lm k}{2}+N\s\label{skn}
\ea 
The distribution for the dominant spin configuration is given by  
\ba
\sj=N(2j+1)\exp\left[-\lm C_j-\s\right]\label{mpd1}					
\ea
For given $k$ and $N$,  $\lm$ and $\s$ are solutions of the following two equations :
\begin{subeqnarray}\label{cons}
\exp [{\s}] &=&\sum_j (2j+1)\exp[-\lm C_j]\slabel{cons1}\\
k/2 &=&N\sum_j C_j(2j+1)\exp[-\lm C_j-\s]\slabel{cons2}
\end{subeqnarray}
where $C_j=\j$. In the appropriate limits \cite{me1,me2} the above two equations reduce to
\begin{subeqnarray}\label{apcons}
e^{\s}&=&\f{2}{\lm^2}\left(1+\f{\sqrt 3}{2}\lm\right)e^{-\f{\sqrt 3}{~2}\lm}\label{siglm}\\
\f{k}{N}&=&1+\f{2}{\lm}+\f{4}{\lm(\sqrt 3\lm+2)}\label{lmro}
\end{subeqnarray}
As far as the thermodynamical aspect  is concerned it is necessary to write the microcanonical entropy in terms of the average area and the number of punctures, first shown in \cite{gp}. Using the relation $k=A_{cl}/4\pi\g\lp^2\simeq\A/4\pi\g\lp^2$ \cite{ae1} the eq.(\ref{skn}) can be cast in the following form
\ba
S_{MC}=\f{\lm\A}{8\pi\g\lp^2}+N\s\label{smc7}
\ea
A graphical analysis in \cite{me1} reveals that for each given value of $k/N>1$ we can find a positive value of $\lm$ from eq.(\ref{lmro}).  Hence,  there exists a unique value of $\g$ given by $\lm/2\pi$ for each given value of $k/N>1$ such that we obtain area term to carry the factor $1/4$. Now, it can be argued on some physical grounds that the term $N\s$ should be a negative definite quantity\cite{me1}, which in turn imposes a bound on $\lm$ given by $1.200<\lm<\infty$.
Hence, from the $\g$-fit it follows that the allowed range of $\g$ is given by $0.191<\g<\infty$. One can look into \cite{me1} for  detailed explanations on this issue.

\subsection{Fixation of the model from investigation of classical limit}
Now, it follows from eq.(\ref{ej}) that the energy eigenvalue of the QIH in a state designated by the spin configuration $\{s_j\}$ will be given by
\ba
\hat H_S|\left\{s_j\right\}\rangle=\lp\sum_j\sum_{n=0}^{\L}a_n(\g)s_jC_j^{~n}|\left\{s_j\right\}\rangle\label{espec}
\ea
One can look into \cite{me2} for an explanation of how $|\{s_j\}\rangle$  forms the eigenstates of the QIH and thus a generic quantum state of the QIH can be written as $|\Psi_S\rangle=\sum_{\left\{s_j\right\}} c_{\left\{s_j\right\}}|\{s_j\}\rangle$
Hence, the expectation value of the Hamiltonian operator or the mean energy for the QIH is given by
\ba
\langle\hat H_S\rangle&=&\langle\Psi_S|\hat H_S|\Psi_S\rangle\nn\\
&=&\lp\sum_{\left\{s_j\right\}}\o[\left\{s_j\right\}]\sum_{j}\sum_{n=0}^{\L}a_n(\g)s_jC_j^{~n}\nn\\
&=&\lp\sum_{j}\sum_{n=0}^{\L}a_n(\g)\sj C_j^{~n}+~\text{sub-dominant contributions }\nn\\
&=&\E\pm\sO(\lp)\label{exe}
\ea
where $\o[\left\{s_j\right\}]=|c[\left\{s_j\right\}]|^2$ is the quantum mechanical probability of the QIH to be found in the state $|\{s_j\}\rangle$. 
Now, we can calculate $\E$ explicitly by using $\sj$ given by eq.(\ref{mpd1}) in the following way 
\ba
\E&\simeq&\lp\sum_{j}\sum_{n=0}^{\L}a_n(\g)\sj C_j^{~n}\nn\\
&=&\lp N\sum_{j}\sum_{n=0}^{\L}a_n(\g) (2j+1)C_j^{~n}\exp(-\lm C_j-\s)\nn\\
&\simeq&\lp N\exp(-\s)\sum_{n=0}^{\L}a_n(\g)\int_{1/2}^{\infty} (2x+1)\[{x(x+1)}\]^{n/2}\exp(-\lm\sqrt{x(x+1)})~dx\nn\\ 
&&~~~~~~~~~~~~~~~~~~~~~~~~~\text{[taking the limit $k\to\infty$ and replacing the sum over $j$ by integration over $x$]}\nn\\
&=&\lp N\exp(-\s)\sum_{n=0}^{\L}a_n(\g)\f{2}{~\lm^{n+2}}\int_{\f{\lm\sqrt{3}}{2}}^{\infty}y^{n+1}e^{-y}dy\nn\\
&&~~~~~~~~~~~~~~~~~~~~~~~~~\text{[applying the change of variable $\lm\sqrt{x(x+1)}=y$ ]}\nn\\
&=&\lp N\exp(-\s)\sum_{n=0}^{\L}a_n(\g)\f{2}{~\lm^{n+2}}~\Gamma\(n+2~,~\lm\sqrt 3/2\)\nn\\
&=& \exp(-\s)F(\L,\lm,\g)\lp N\label{estar}
\ea
where $F(\L,\lm,\g)=\sum_{n=0}^{\L}a_n(\g)\f{2}{~\lm^{n+2}}~\Gamma\(n+2~,~\lm\sqrt 3/2\)$,
which is obviously  a positive definite function of $\lm$. 
Similar to the above calculation one can also derive the expression for the mean area as follows 
\ba
\A&\simeq&8\pi\g\lp^2\sum_{j}\sj C_j\nn\\
&=&8\pi\g\lp^2 N\sum_{j} (2j+1)C_j\exp(-\lm C_j-\s)\nn\\
&\simeq&8\pi\g\lp^2 N\exp(-\s)\int_{1/2}^{\infty} (2x+1)\[{x(x+1)}\]^{1/2}\exp(-\lm\sqrt{x(x+1)})~dx\nn\\ 
&&~~~~~~~~~~~~~~~~~~~~~~~~~\text{[taking the limit $k\to\infty$ and replacing the sum over $j$ by integration over $x$.]}\nn\\
&=&8\pi\g\lp^2 N\exp(-\s)\f{2}{~\lm^{3}}\int_{\f{\lm\sqrt{3}}{2}}^{\infty}y^{2}e^{-y}dy\nn\\
&&~~~~~~~~~~~~~~~~~~~~~~~~~\text{[applying the change of variable $\lm\sqrt{x(x+1)}=y$ ]}\nn\\
&=&\exp(-\s)\f{16\pi\g}{~\lm^{3}}~\Gamma\(3,\lm\sqrt 3/2\)\lp^2 N\label{astar}
\ea
Collecting eq.(\ref{estar}) and eq.(\ref{astar}) we have a relation between the mean values of energy and area of the QIH given by
\ba
\E&=&\f{\lm^3}{16\pi\g\lp}\f{F(\L,\lm,\g)}{\Gamma(3,\lm\sqrt 3/2)}~\A\nn\\
&=&\f{\xi(\L,\lm,\g)}{\lp}~\A\label{ea}
\ea
where the quantity $\xi(\L,\lm,\g)=\f{\lm^3}{16\pi\g}\f{F(\L,\lm,\g)}{\Gamma(3,\lm\sqrt 3/2)}$ can be explicitly written as
\ba
\xi(\L,\lm,\g) =\f{\lm^3}{16\pi\g\Gamma(3,\lm\sqrt 3/2)}\sum_{n=0}^{\L}a_n(\g)\f{2}{~\lm^{n+2}}~\Gamma\(n+2~,~\lm\sqrt 3/2\)\label{xi}
\ea
Now, let us recollect that the $\g$-fit is required only to manifest the BHAL. Otherwise $S_{MC}$ will be given by eq.(\ref{smc7}), the Lagrange multipliers being functions of $k$ and $N$. So, we shall always make the $\g$-fit after performing all the calculations. The microcanonical entropy {\it without the $\g$-fit}, given by eq.(\ref{smc7}) can be expressed in terms of the equilibrium energy of the QIH using eq.(\ref{ea}) as
\ba
S_{MC}=\f{\lm}{8\pi\g\xi\lp}\E+N\s
\ea
Now, on identifying the above expression with the usual thermodynamic form of the microcanonical entropy given by the expression $S_{MC}=\b\E+N\s$,
 the form of the equilibrium temperature $(T=1/\beta)$ is quite clearly manifest, which, using eq.(\ref{xi}), can be written as 
\ba
T&=&\f{8\pi\g\xi\lp}{\lm}\nn\\
&=&\f{8\pi\g\lp}{\lm}\f{\lm^3}{16\pi\g\Gamma(3,\lm\sqrt 3/2)}\sum_{n=0}^{\L}a_n(\g)\f{2}{~\lm^{n+2}}~\Gamma\(n+2~,~\lm\sqrt 3/2\)\nn\\
&=&\f{\lp}{\Gamma(3,\lm\sqrt 3/2)}\sum_{n=0}^{\L}\f{a_n(\g)}{~\lm^{n}}~\Gamma\(n+2~,~\lm\sqrt 3/2\)\label{temp}
\ea
Considering $\g=\lm/2\pi$, the above expression for temperature becomes
\ba
T&=&\f{\lp}{\Gamma(3,\sqrt 3\pi\g)}\sum_{n=0}^{\L}\f{a_n(\g)}{~(2\pi\g)^{n}}~\Gamma\(n+2~,~\sqrt 3\pi\g\)\label{tempgamma}
\ea
It is worthy of mentioning for the sake of clarity that  the first law is given by $T \d S=\d E+\mu \d N$, where $\mu=-\sigma/T$ is the `chemical potential' corresponding to the `quantum hair' $N$\cite{gp}. Just for notational familiarity, the suffixes, etc. has been dropped.

\subsection{Single Puncture Energy Spectrum as a power law}\label{sec3.1}
As far as this work is concerned we have only tried to construct a model energy spectrum for the QIH which is {\it not} the result of a true quantization of the horizon energy. So, we shall try to look at some of the consequences of this model energy spectrum by making some specific choices. Let us investigate the special case where the single puncture energy spectrum follows a power law of the single puncture area contribution i.e. we choose $b_m=\n_n\d_n^m\lp^{-2n}$ with $m\geq1$. The single puncture energy spectrum given by eq.(\ref{ej}) written as a polynomial  gets reduced to a power law given by
\ba
E_j=\lp b_mA_j^{~m}=\lp \underbrace{\n_m(8\pi\g)^m}_{a_m(\g)}C_j^{~m}\label{ejpl}
\ea
where $\n_m$ is an unknown positive constant.
As a consequence of the power law, the expression of the temperature given by eq.(\ref{temp}) is now reduced to
\ba
T&=&\f{\lp}{\Gamma(3,\lm\sqrt 3/2)}\f{\n_n(8\pi\g)^n}{~\lm^{n}}~\Gamma\(n+2~,~\lm\sqrt 3/2\)\label{tempo}
\ea
For the BHAL to be valid, we must have $\g=\lm/2\pi$ which reduces eq.(\ref{tempo}) to
\ba
T&=&\f{\lp}{\Gamma(3,\sqrt 3\pi\g)}\n_n4^n~\Gamma\(n+2~,~\sqrt 3\pi\g\) \label{sintemp}
\ea
Hence, the temperature clearly carries the exponent of the power law which governs the single puncture energy spectrum and that is quite disturbing as far as our knowledge of usual thermodynamics is concerned. But, since we have proposed a model energy spectrum, we shall demand that this model must consistently yield the known results of the classical IH thermodynamics by choosing the coefficients $\n_m$ appropriately.
\par
{\it Choosing the coefficients :} Just for the moment if we go back to the general single puncture energy spectrum given by eq.(\ref{ej}), one can argue that the coefficients must decrease with increasing $m$ so as to make the energy spectrum convergent. But, most importantly, we must get rid of the parameter governing the single puncture energy spectrum from appearing in the expression of the temperature in eq.(\ref{sintemp}). Hence, from this power law analysis we have a good reason to choose $\n_m=\n\Gamma(3,\sqrt 3\pi\g)/ 4^{m}\Gamma\(n+2~,~\sqrt 3\pi\g\)$, $\n$ is a scaling constant of the energy spectrum.  As a result the temperature in eq.(\ref{tempo}) is now given by
\ba
T= \n\lp\left[\f{2\pi\g}{\lm}\right]^m\label{tempo1}
\ea
which, for the BHAL to be valid i.e. for $\g=\lm/2\pi$, reduces to 
\ba
T=\n\lp
\ea
That the choice of the coefficients is appropriate can be understood from the fact that the temperature is now independent of the single puncture energy spectrum of the QIH. 

\subsection{Revisiting the Complete Energy Spectrum}\label{sec3.2}
Having done all these analyses in the previous sections, now we at least have got a hint about the coefficients $a_n$ so as to get an explicit structure of the full energy spectrum of the QIH. Since we only proposed the single puncture energy contribution given by eq.(\ref{ej}) which resulted in the QIH energy spectrum in eq.(\ref{espec}), we did not have any idea about the coefficients $a_n$. But, considering the choice of the coefficients $a_n$ from the power law analysis shown above and taking into account the validity of the BHAL ($\g=\lm/2\pi$), we can propose the single puncture energy spectrum to be 
\ba
E_j=\n\Gamma(3,\sqrt 3\pi\g)\lp\sum_{n=0}^{\L}\f{(2\pi\g)^n}{\Gamma\(n+2~,~\sqrt 3\pi\g\)}~C_j^{~n}\label{ej1}
\ea
from which it follows that the energy spectrum for a QIH will be given by
\ba
\hat H_S|\left\{s_j\right\}\rangle
&=&\n\Gamma(3,\sqrt 3\pi\g)\lp\sum_j\sum_{n=0}^{\L}\f{(2\pi\g)^n}{\Gamma\(n+2~,~\sqrt 3\pi\g\)}~s_jC_j^{~n}|\left\{s_j\right\}\rangle\label{espec1}
\ea
 The form of the coefficients can be explicitly written as 
\ba
a_n(\g)=\f{\n\Gamma(3,\sqrt 3\pi\g)(2\pi\g)^n}{\Gamma\(n+2~,~\sqrt 3\pi\g\)}
\ea 
where $\n$ is some unknown positive constant. The other relevant quantities can be written down as
\ba
F(\L,\g)&=&\f{\Gamma(3,\sqrt 3\pi\g)}{2\pi^2\g^2}\eta(1+\L)\nn\\
\xi(\L,\g)&=&\f{1}{4}\eta(1+\L)\nn
\ea
The expression of the equilibrium temperature given by eq.(\ref{tempgamma}) reduces to
\ba
T=\n(1+\L)\lp\label{tempL}
\ea
It is evident from the expression for the temperature given by eq.(\ref{tempL}) that the cut-off imposed on the sum over $n$ is justified because the temperature diverges for $\L\to\infty$. Hence, at this moment we can not model the single puncture energy spectrum with some series with infinite terms and the cut off $(\L)$ seems to be absolutely necessary. Now, as we are investigating a model Hamiltonian, we shall expect to yield the results already known by choosing the parameters suitably. 
Since we have worked with fixed $N$ and hence a non zero $\sigma$, we shall refer to the quasi local first law discusses in \cite{gp,gp2}. In \cite{gp,gp2} it has been somehow argued that the local observers who remains stationary with respect to the horizon \cite{gp2} will observe a universal surface gravity. We fix our model by choosing $\n=\L/(1+\L)$ so that eq.(\ref{tempL}) reduces to
\ba
T=\L\lp\label{tempfix}
\ea
where $\L$ is the constant temperature to be observed by the local observer.

\subsection{Spectrum of the Hamiltonian operator}
Hence, the most general structure possible for the Hamiltonian operator associated with the QIH, as observed by the local observer is given by
\ba
\hat H_S|\left\{s_j\right\}\rangle
&=&\f{\L\Gamma(3,\sqrt 3\pi\g)\lp}{(1+\L)}\sum_j\sum_{n=0}^{\L}\f{(2\pi\g)^n}{\Gamma\(n+2~,~\sqrt 3\pi\g\)}~s_jC_j^{~n}|\left\{s_j\right\}\rangle\label{espec1}
\ea
It may be mentioned that the fixation of $\eta$ and $\L$ will be somewhat different if we consider the first of thermodynamics associated with the IH as derived in the classical theory \cite{ih3}. The situation is physically quite different from the current scenario, has several different implications and hence, needs to be discussed separately.

\section{Discussion}\label{sec5}


The thermodynamics associated with the kind of energy spectrum of a QIH considered here, has never been studied earlier in literature. This is due to the complete different viewpoint of going from quantum to classical regime presented in \cite{hamqih} motivated by some deep underlying issues regarding the theories of IH and QIH.   Generally, one considers the horizon energy as a function of the horizon area (e.g. power law). This intuition works in our mind due to our instinctive affinity to look at a quantum theory through the classical spectacles. To be more explicit, in numerous cases of the study of black holes the mass formula for known black hole solutions are expressed in terms of the area and addressed as the mass spectrum of the black hole \cite{stab1,stab2,stab3,stab4,stab5,stab6,stab7}. In fact, many a times, in such formulae, the area spectrum of LQG is used directly in the classical formula and the mass of the black hole is considered to be quantized \cite{stab2,stab4,cr} which is of course not a true quantization of the horizon energy and also devoid of any physical justification, apart from being an ad hoc assumption. A genuine energy spectrum for a QIH should be derived by quantization of the classical notion of horizon energy similar to the quantization of area, volume and length resulting in the corresponding operators in quantum gravity \cite{op1,op2,op3,op4,op5}. The other alternative is to propose one, based on solid physical arguments, which has been done in the companion paper\cite{hamqih}. Interestingly, the result provided by our model energy spectrum in eq.(\ref{ea}) from a purely quantum statistical and thermodynamical perspective, is capable of explaining the results of \cite{eb1,eb2,gp2,smo} or \cite{ih3}. It remains to be seen whether it is possible to construct the corresponding Hamiltonian operator for the QIH yielding the kind of energy spectrum discussed here from the classical theory, where of course $\L$ will be known a priori. 
\par
Now, as far as the thermodynamic aspect of this paper is concerned, it is worth mentioning that unlike \cite{ae1,ae2,ae3} we  deal with usual canonical {\it energy} ensemble as we have an explicit structure of the Hamiltonian. We need not use a `Boltzmann-like' factor $e^{-\alpha A}$ in the canonical partition function\cite{ae1}, accompanied by a fictitious conjugate parameter $\alpha$, alongside the Boltzmann factor $e^{-\b E}$. As far as \cite{ae2,ae3} are concerned, it is a pure area ensemble involving only $e^{-\alpha A}$ in the canonical partition function and devoid of the Boltzmann factor $e^{-\b E}$. All of these approaches are significant and interesting by their own virtue. But none of them actually attacks the problem of black hole horizon thermodynamics following usual canonical energy ensemble due to the lack of knowledge of the Hamiltonian and the energy spectrum associated with the horizon. This is where the use of the proposed model Hamiltonian reap the benefits and allow us to follow usual canonical {\it energy} ensemble approach to thermodynamic analysis of QIHs .

\vspace{1cm}
{\bf Acknowledgment :} I am grateful to Deepak Vaid for offering comments which helped me to improve and correct some parts of the paper. I also want to thank the referee of this paper for offering valuable suggestions which helped me improving on an earlier version of this paper. I am thankful to the Department of Science and Technology of India for providing financial support to carry out my research.

\end{document}